\documentclass[a4paper,11pt]{article}
\usepackage{rotate}
\usepackage{epsfig}
\usepackage{rkuhn}
\begin{document}
\setlength{\baselineskip} {2.5ex}
\def\Journal#1#2#3#4{{#1} {\bf #2}, #3 (#4)}
\def\NCA{\em Nuovo Cimento}
\def\NIM{\em Nucl. Instr. Meth.}
\def\IJMPA{{\em Int. Jour. Mod. Phys.} A}
\def\NIMA{{\em Nucl. Instr. Meth.} A}
\def\NPA{{\em Nucl. Phys.} A}
\def\NPB{{\em Nucl. Phys.} B}
\def\NPP{\em Nucl. Part. Phys.}
\def\PLB{{\em Phys. Lett.} B}
\def\SJPN{\em Sov. Jour. Part. Nucl.}
\def\SJNP{\em Sov. Jour. Nucl. Phys.}
\def\PRL{\em Phys. Rev. Lett.}
\def\PR {\em Phys. Rev.} 
\def\PRC{{\em Phys. Rev.} C}
\def\PRD{{\em Phys. Rev.} D}
\def\ZPC{{\em Z. Phys.} C}

\begin{center} 
{Contribution to Advanced Study Institute, "Symmetries and Spin"
- Praha-SPIN-2002,\\ Workshop Chairman, M. Finger,
http://mfinger.home.cern.ch/mfinger/praha2002/
\\ Prague, Czech Republic, July 2002}\\    
\vspace{0.1cm} 
{\Large\bf Pion and Kaon Polarizabilities at CERN COMPASS\\}
\vspace{0.1cm} 
\textsc{\large Murray Moinester
\footnote{ together with: F. Balestra, R. Bertini, M.P.
Bussa, M. Colantoni, O. Denisov, A. Dolgopolov, M. Faessler, A. Ferrero,
L. Ferrero, J. Friedrich, V. Frolov, R. Garfagnini, N.
Grasso, V. Kolossov, R. Kuhn, A. Maggiora, M. Maggiora, A.
Manara, Y. Mikhailov, V. Obraztsov, A.
Olchevski, D. Panzieri, S. Paul, G, Piragino, J.
Pochodzalla, V. Poliakov, A. Sadovski, M. Sans, L.
Schmitt, H. Siebert, A. Skachkova, T. Walcher, A.
Zvyagin} 
\\ for the COMPASS collaboration}\\
\textit{R. and B. Sackler Faculty of Exact Sciences,\\ 
School of Physics and Astronomy, Tel
Aviv University,\\ 69978 Tel Aviv, Israel\\ 
murraym@tauphy.tau.ac.il\\}
\end{center}

\begin{center}
Abstract: 
\end{center}


{\bf Objective:}
  
The electric $(\bar{\alpha})$ and magnetic $(\bar{\beta})$ pion Compton
polarizabilities characterize the pion's deformation in the
electromagnetic field of the $\gamma$ during $\gamma\pi$ Compton
scattering.  They depend on the rigidity of the pion's internal structure
as a composite particle. The polarizabilities deduced by Antipov et al.
in their low statistics Primakoff experiment ($\sim$ 7000 events) were
$\bar{\alpha}_{\pi} = -\bar{\beta}_{\pi} = 6.8 \pm 1.4 \pm 1.2$, in units
of $10^{-43}$ cm$^3$.  This value, ignoring the large error bars, is
about three times larger than the chiral perturbation theory ($\chi$PT)
prediction.  Taking into account the very high beam intensity, fast data
acquisition, high acceptance and good resolution of the CERN COMPASS
experiment, one can expect from COMPASS statistics a factor 6000 higher,
a data sample that includes many tests to control systematic errors, and
a significantly reduced total measurement uncertainty for
$\bar{\alpha}_{\pi}$, of order 0.4.

{\bf Methodology:} 

CERN COMPASS studies of pion-photon
interactions, to achieve a unique Primakoff physics program centered on
pion polarizability studies. We use 100-200 GeV pion beams and a virtual
photon target, and magnetic spectrometers and calorimeters to measure the
complete kinematics of pion-photon reactions.  COMPASS was set up during
2000/01, including a successful Primakoff test run, and then began data
taking with a muon beam for the proton spin physics component of its
program. COMPASS will next run its spin physics program and Primakoff
program preparations, followed by its pion beam physics program,
including pion polarizability.  For pion polarizability, $\gamma\pi$
scattering will be measured via radiative pion scattering (pion
Bremsstrahlung) in the nuclear Coulomb field:  $\pi + Z \rightarrow \pi'
+ \gamma + Z.$ A virtual photon from the Coulomb field of the target
nucleus is scattered from the pion and emerges as a real photon
accompanying the pion at small forward angles in the laboratory frame,
while the target nucleus (in the ground state)  recoils with a small
transverse momentum kick p$_t$.  The radiative pion scattering reaction
is equivalent to $\gamma$ + $\pi$ $\rightarrow$ $\gamma$ + $\pi$
scattering for laboratory $\gamma$'s of order 1 GeV incident on a target
$\pi$ at rest.  The pion polarizabilities are determined by their effect
on the shape of the measured $\gamma \pi$ Compton scattering angular
distribution.

{\bf Significance:}  

The pion polarizabilities are key observables, and provide stringent
tests of our understanding of chiral symmetry, its spontaneous breakdown,
the role of explicit symmetry breaking in QCD, and consequently the very
foundations of nuclear physics.  The $\chi$PT effective Lagrangian, using
data from radiative pion beta decay, predicts the pion electric and
magnetic polarizabilities $\bar{\alpha}_{\pi}$ = -$\bar{\beta}_{\pi}$ =
2.7 $\pm$ 0.4. New high precision pion polarizability measurements via
radiative pion scattering data from COMPASS will provide important new
tests of this QCD chiral dynamics prediction.

\newpage
\section{Scientific Background:} 

Pion polarizabilities will be measured at the CERN COMPASS experiment
\cite{paul,cd,bormio,tm,sans,kuhn,mlc,of}, a new high priority approved
spectrometer facility at CERN that uses muon and pion beams for studies of hadron
structure and spectroscopy.  The polarizabilities are obtained from measurements
of the $\gamma \pi \rightarrow \gamma \pi$ gamma-pion Compton scattering. For the
pion, chiral perturbation theory ($\chi$PT) leads to precision predictions for the
polarizabilities \cite{hols1,buergi,babu2,tau}.  
Precision measurements of
polarizabilities therefore subject the $\chi$PT techniques of QCD
to new and serious tests. The polarizability measurements described here 
are a part of the global COMPASS Primakoff program \cite 
{cd,bormio,tm}
to study 
pion and Kaon polarizabilities, chiral anomalies, and pionic and kaonic
Hybrid mesons. The global COMPASS physics program was described Sept. 2002 
at a CERN COMPASS Future workshop \cite {future}.

\subsection{Pion Polarizabilities via Primakoff Scattering}

\indent
For the pion polarizability, $\gamma\pi$ scattering was measured (with large
uncertainties)  with 40 GeV pions \cite{anti1} via radiative pion scattering
(pion Bremsstrahlung) in the nuclear Coulomb field:
\begin{equation}
\label{eq:polariz}
\pi + Z \rightarrow \pi' + \gamma + Z'.
\end{equation}

In this measurement, the incident pion Compton scatters from a virtual photon in
the Coulomb field of a nucleus of atomic number Z; and the final state $\gamma$
and pion are detected in coincidence.  The radiative pion scattering reaction is
equivalent to  $\gamma$ + $\pi^{-}$ $\rightarrow$  $\gamma$ + $\pi^{-}$
scattering for laboratory $\gamma$'s of order 1 GeV incident on a target
$\pi^{-}$ at rest. It is an example of the well tested Primakoff formalism
\cite{jens,ziel} that relates processes involving real photon interactions to
production cross sections involving the exchange of virtual photons.

In the 40 GeV radiative pion scattering experiments, it was shown experimentally
\cite{anti1} and theoretically \cite{galp} that the Coulomb amplitude clearly
dominates, and yields sharp peaks in t-distributions at very small squared four
momentum transfers (t) to the target nucleus t $\leq 6 \times 10^{-4}$
(GeV/c)$^{2}$. Backgrounds from strong processes were low, and are expected to be
even lower at the higher energy ($\sim$ 190 GeV) planned for the CERN COMPASS
experiment.

{\bf All polarizabilities in this paper are expressed in units of $10^{-43}$
cm$^3$}. The $\chi$PT 1-loop prediction \cite {hols1,buergi} for the pion
polarizability is $\bar{\alpha}_{\pi}=~-\bar{\beta}_{\pi} = 2.7 \pm ~0.4$; with
values $\bar{\alpha}_{\pi}= 2.4 \pm ~0.5;~\bar{\beta}_{\pi} = -2.1 \pm ~0.5$ at
two-loop \cite {buergi}.  Holstein \cite {hols1} showed that meson exchange via a
pole diagram involving the a$_1$(1260) resonance provides the main contribution
($\bar{\alpha}_{\pi}$ = 2.6) to the polarizability.  Xiong, Shuryak, Brown (XSB)  
\cite{xsb} assuming a$_1$ dominance find $\bar{\alpha}_{\pi}$ = 1.8.  Many other
QCD based polarizability calculations, including lattice QCD \cite{lattice},  are 
also available, 
as described in
different polarizability articles \cite {cd,babu2,Pennington}. 
In this report, we emphasize comparisons with
model independent chiral perturbation predictions.
But we will of course compare the data to all theoretical predictions.

For the kaon, the lowest order $\chi$PT prediction \cite {cd,hols1,pol} 
is $\bar{\alpha}_{K^-}$ = 0.5 .
The kaon polarizability measurements at COMPASS should complement those for pion
polarizabilities for chiral symmetry tests away from the chiral limit. More
extensive studies of kaon polarizabilities were given in Ref.  
\cite {ev,gp}. Until now, only an upper limit \cite {gb} at 90\% confidence was
measured (via energy shifts in heavy Z kaonic atoms) for the $K^-$, with
$\bar{\alpha}_{K} \leq 200.$ Kaon polarizability measurements have never been
carried out. For the kaon
polarizability, due to the lower beam intensity, the statistics will be roughly 50
times less than for the pion case. But COMPASS would still obtain the first 
ever Kaon
polarizability measurement. 

\subsection{Pion Polarizabilities}

For the $\gamma$-$\pi$ interaction at low energy, chiral perturbation theory
($\chi$PT) provides a rigorous way to make predictions; because it stems
directly from QCD and relies only on the solid assumptions of spontaneously
broken SU(3)$_L$ $\times$ SU(3)$_R$ chiral symmetry, Lorentz invariance and
low momentum transfer. Unitarity is achieved by adding pion loop corrections
to lowest order, and the resulting infinite divergences are absorbed into
physical (renormalized) coupling constants L$^r_i$ (tree-level coefficients
in L$^{(4)}$, see Refs. \cite{gass1,donn1}). With a perturbative expansion
of the effective Lagrangian limited to terms quartic in the momenta and
quark masses (O(p$^4$)), the method establishes relationships between
different processes in terms of the L$^r_i$. For example, the radiative pion
beta decay and electric pion polarizability are expressed as \cite{donn1}:
\begin{equation} h_A/h_V = 32\pi^2(L^r_9+L^r_{10}); \bar{\alpha}_{\pi} =
\frac{4\alpha_f}{m_{\pi}F^{2}_{\pi}}(L^r_9+L^r_{10}); \label{eq:L9}
\end{equation} \noindent where F$_\pi$ = 93.1 MeV \cite{part} is the pion
decay constant, h$_A$ and h$_V$ are the axial vector and vector coupling
constants in the decay, and $\alpha_f$ is the fine structure constant. The
experimental ratio \cite{part} h$_A$/h$_V$ = 0.45 $\pm$ 0.06, leads to
$\bar{\alpha}_{\pi}$ = -$\bar{\beta}_{\pi}$ = 2.7 $\pm$ 0.4, where the error
shown is due to the uncertainty in the h$_A$/h$_V$ measurement \cite{babu2,tau}.
For the Kaon polarizability, Eq. 2 is used with the Kaon mass instead of the 
pion mass.  

The pion polarizabilities deduced by Antipov et al. \cite{anti1} in their low
statistics experiment ($\sim$ 7000 events) were $\bar{\alpha}_{\pi} =
-\bar{\beta}_{\pi} = 6.8 \pm 1.4 \pm 1.2$, with the analysis constraint that
$\bar{\alpha}_{\pi} + \bar{\beta}_{\pi} = 0$, as expected theoretically \cite
{hols1}.  The deduced polarizability value, not counting the large error bars, is
some three times larger than the $\chi$PT prediction. {\bf The available
polarizability results have large uncertainties. There is a clear need for new and
improved radiative pion scattering data.}

\subsection{Meson Radiative Transitions}

COMPASS will also study Primakoff radiative transitions leading from the pion to
the $\rho^-$, a$_1$(1260), and a$_2$(1320); and for the kaon to K$^*$. The $\rho$
data is obtained with the $\gamma\pi$ polarizability trigger, while the others
require a particle multiplicity trigger \cite{tm}, as discussed later in Section 
3.3.
Radiative transition widths are
predicted by vector dominance and quark models. Independent and higher precision
data for these and higher resonances would be valuable as a check of the 
COMPASS apparatus
and in order to allow a more meaningful comparison with theoretical predictions.
For example, the $\rho \rightarrow \pi \gamma$ width measurements
\cite{jens,hust,capr} range from 60 to 81 keV; the a$_1$(1260) $\rightarrow \pi
\gamma$ width measurement (\cite{ziel}) is $0.64 \pm 0.25 $ MeV; and the
a$_2$(1320) $\rightarrow \pi \gamma$ width is is $\Gamma = 295 \pm$ 60 keV
\cite{ciha} and $\Gamma = 284 \pm 25 \pm 25$ keV \cite{mol}. For K$^* \rightarrow
K \gamma$, the widths obtained previously are 48 $\pm$ 11 keV \cite {berg} and 51
$\pm$ 5 keV \cite {chan}.

According to Holstein \cite {hols1} and XSB \cite {xsb}, the a$_1(1260)$ width and
the pion polarizability are related to one another.  For a$_1$(1260) $\rightarrow
\pi \gamma$, the experimental width \cite {ziel} is $\Gamma = 0.64 \pm 0.25$ MeV.  
XSB \cite {xsb} used an estimated radiative width $\Gamma$ = 1.4 MeV, higher than
the experimental value \cite {ziel}, as determined in the Primakoff reaction $\pi
Z \rightarrow a_1 Z$, followed by $a_1^- \rightarrow \pi^- \rho$.  It is with this
estimated width that they calculate the pion polarizability to be
$\bar{\alpha}_{\pi}$ = 1.8. COMPASS will experimentally check the a$_1$
assumptions of XSB via the Primakoff reaction $\pi Z
\rightarrow a_1 Z$, and the consistency of the expected relationship of this
radiative width and the pion polarizability \cite {qgp}.
 
\section{Research Goals and Expected Significance:}

We studied the statistics attainable and uncertainties achievable for the pion
polarizabilities in the COMPASS experiment, based on Monte Carlo simulations. We
begin with an estimated $\sigma(Pb)=0.5 mb$ Compton scattering cross section per
Pb nucleus and a total inelastic cross section per Pb nucleus of 0.8 barn.  High
statistics will allow systematic studies, with fits carried out for different
regions of photon energy $\omega$, Z$^2$, etc.;  and polarizability determinations
with statistical uncertainties lower than 0.1.

 We consider a pion beam flux of $2 \times 10^{7}$ pions/sec, with 
a spill
structure that provides a 5 second beam every 16 seconds.  For pion 
polarizability, in
2 months of running at 100\% efficiency, we obtain 3.2$\times$ 10$^{13}$ beam
pions. We use a 0.8~\% interaction length target, 3 mm Lead plate with target
density $N_t=10^{22} cm^{-2}$. The Primakoff interaction rate is then $R =
\sigma(Pb) \cdot N_t= 5. \times$ 10$^{-6}$. Therefore, in a 2 month run, one
obtains 1.6 $\times$ 10$^{8}$ Primakoff polarizability events at 100\% efficiency.  
Considering efficiencies for tracking (92\%), $\gamma$ detection (58\%),
accelerator and COMPASS operation (60\%), analysis cuts to reduce backgrounds
(75\%), we estimate a global efficiency of $\epsilon$(total)=24\%, or 4.$\times$
10$^7$ useful pion polarizability events per 2 month run. Prior to the data
production run, time is also needed to calibrate ECAL2, to make the tracking
detectors operational, to bring the DAQ to a stable mode, and for other
contingencies. The above expected statistics in a two month data production run is
a factor 6000 higher than the 7000 events of the previous pion polarizability
Primakoff experiment.

We will also access kaon polarizabilities considering the approximately 2\% kaon
component of the beam. We will use the CERN CEDAR Cherenkov beam detector for the
kaon particle identification. Statistics of order $4. \times 10^5$ events would
allow a first time determination of the kaon polarizability.

COMPASS provides a unique opportunity to investigate pion and Kaon
polarizabilities.  Taking into account the very high beam intensity, fast data
acquisition, high acceptance and good resolution of the COMPASS setup, one can
expect from COMPASS the highest statistics and a `systematics-free' data sample
that includes many tests to control possible systematic errors.

\section{Detailed description of Research Program:}

COMPASS is a fixed target experiment which runs primarily with a 160 GeV
polarized muon beam and a 190 GeV pion beam.
In order to achieve a good energy resolution within a wide energy range,
COMPASS is designed as a two stage spectrometer with 1.0 Tm and 5.2 Tm
conventional magnets. The tracking stations are composed of different
detector
types to cover a large area while achieving a good spatial resolution in the
vicinity of the beam. Most of the tracking detectors operate on the principle
of gas amplification, while some are silicon strip detectors. At the end of each
stage, an electromagnetic and a
hadronic calorimeter detects the energies of the gammas, electrons
and hadrons.  The calorimeters of the first stage and the EM calorimeter of
the second stage have holes through which the beam passes. 

We considered elsewhere in detail the beam, detector, target, and trigger 
requirements for
polarizability studies in the CERN COMPASS experiment \cite
{cd,bormio,tm}. We describe these studies briefly below.
The Sept. 2002 status of the COMPASS apparatus is described in Ref. \cite 
{future}. 

\subsection{Monte Carlo Simulations}

The setup we used for the Monte Carlo simulations was the official setup for the
year 2001 run with the addition of three GEM stations and six silicon stations as
projected for the year 2002 run. The additional detectors allow more precise
tracking. We carried out the polarizability simulations using (1) the POLARIS
event generator with (2) the CERN COMPASS GEANT (COMGEANT) package \cite {va},
whose output is a ZEBRA file with the information on the traces left by particles
in detectors, (3) the \coral\ \compass\ reconstruction and analysis library,
structured as a set of modules:  an input package is used to read the ZEBRA files
produced by \comgeant, \traffic (TRAck Finding and FItting in Compass),
calorimeter and RICH packages, a ROOT output package, the detector data decoding
package, etc., and (4) the new CERN histograming and display and fit program ROOT.  
We use the terms \emph{generated} and \emph{reconstructed}, the first denoting the
input physics events to \comgeant\ from POLARIS and the latter the output of
\coral\ which contains the reconstruction of these events in the COMPASS
spectrometer.

\begin{figure}[tbc]
\centerline{\epsfig{file=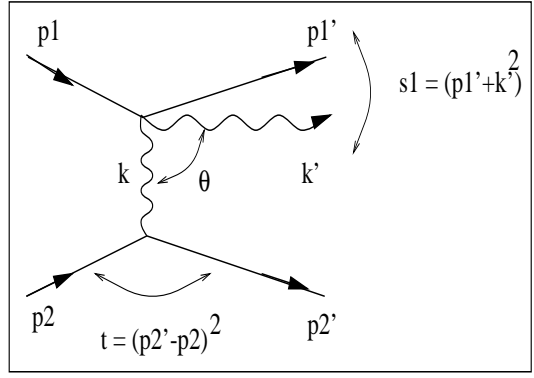,width=7cm,
height=5cm}}
\caption{The Primakoff $\gamma$-hadron Compton process and kinematic variables
(4-momenta): p1, p1$^\prime$ = for initial/final hadron, p2, p2$^\prime$ = for 
initial/final target, k, k$^\prime$ = for initial/final gamma, and $\theta$ 
the scattering angle of the $\gamma$ in the alab frame.}
\label{fig:diagram}
\end{figure}

POLARIS produces events of the type Eq.~\ref{eq:polariz}, based on the theoretical
Primakoff $\gamma\pi$ Compton scattering cross section. The four-momentum of each
particle is p1, p2, p1$^\prime$, p2$^\prime$, k, k$^\prime$, respectively, as
shown in Fig.~\ref{fig:diagram}. In the one-photon exchange domain, this reaction
is equivalent to $\gamma + \pi \rightarrow \gamma^\prime + \pi^\prime$, and the
four-momentum of the incident virtual photon is k = q = p2$-$p2$^\prime$. We have
therefore t=k$^2$ with t the square of the four-momentum transfer to the nucleus,
F(t) the nuclear form factor (essentially unity at small t, $\sqrt{\rm{s}}$ the
mass of the $\gamma\pi$ final state, and t$_0$ the minimum value of t to produce a
mass $\sqrt{\rm{s}}$. The momentum modulus $|\vec{k}|$ (essentially equal to
p$_T$) of the virtual photon is in the transverse direction, and is equal and
opposite to the momentum p$_T$ transferred to the target nucleus.  The pion
polarizability is extracted via a fit of the theoretical cross section to the
scattered $\gamma$ angular distribution in the projectile (alab) rest frame. The
total Primakoff cross section is computed by integrating numerically the
differential cross section $\sigma(s,t,\theta)$ of Eq.~\ref{eq:Primakoff_1} below
for the Primakoff Compton process.

\subsection{Primakoff $\gamma\pi$ Compton Event Generator}

We describe the event generator for the radiative scattering of the pion in the
Coulomb field of a nucleus \cite{pol}. In the pion alab frame, the nuclear Coulomb
field of target M$_Z$ effectively provides a virtual photon beam incident on a
pion target at rest. We have for the variable t=k$^2 =q^2 \equiv$ M$^2$, where t
is the 4-momentum transferred to the nucleus, and M is the virtual photon mass.
Since t=2M$_Z$[M$_Z$-E(Z',lab)]$<$0, the virtual photon mass is imaginary. To
approximate real pion Compton scattering, the virtual photon should be taken to be
almost real. For small t, the electromagnetic contribution to the scattering
amplitude is large compared to meson and Pomeron (diffractive) exchange
contributions. Radiative corrections for Primakoff scattering have been calculated
to be very small \cite {ak}.

The Primakoff differential cross section of the process of Eq.~\ref{eq:polariz} in
the alab frame may be expressed as \cite{starkov}:
\begin{equation}
\label{eq:Primakoff_1}
\frac
{{d}^3{\sigma}}
{{dt}{d}{\omega}{d\cos{\theta}}}
=
\frac
{\alpha_{f}{Z}^2}
{\pi\omega}
\cdot
\frac
{t-t_{0
}}
{t^2}
\cdot
\frac
{{d}\sigma_{\gamma\pi}{(}\omega,\theta{)}}
{{d}{\cos}{\theta}}\cdot F^2(t),
\end{equation}
where the $\gamma\pi$ cross section is given by:
\begin{equation}
\label{eq:Primakoff_2}
\frac
{{d}\sigma_{\gamma\pi}{(}\omega,\theta{)}}
{{d}{\cos}{\theta}}
=
\frac
{{2}{\pi}{\alpha_{f}}^2}
{{m}_{\pi}^2}
\cdot
\{
{F}_{\gamma\pi}^{pt}{(}{\theta}{)}
+
\frac
{{m_{\pi}}{\omega}^2}
{\alpha_{f}}
\cdot
\frac
{\bar{\alpha}_{\pi}{(}1+{\cos}^{2}{\theta}{)}+2\bar{\beta}_{\pi}{\cos\theta}}
{{(}{1+\frac{\omega}{m_{\pi}}{(1-\cos{\theta})}}{)}^3}
\}.
\end{equation}
Here, t$_0$=$(m_{\pi}\omega/p_{b})^2$, with $p_{b}$ the incident pion beam
momentum in the laboratory, $\theta$ the scattering angle of the real photon
relative to the incident virtual photon direction in the alab frame, $\omega$ the
energy of the virtual photon in the alab frame, $Z$ the nuclear charge, $m_\pi$
the pion mass, $\alpha_{f}$ the fine structure constant, 
F(t) is the nuclear electromagnetic form factor (approximately unity in the
range $t < 2.5 \times 10^{-4} GeV^2$), 
and $\bar{\alpha_\pi}$,
$\bar{\beta_\pi}$ the pion electric and magnetic polarizabilities. The energy of
the incident virtual photon in the alab (pion rest) frame is:
\begin{equation}
\omega \sim  (s - {m_{\pi}}^2)/2m_{\pi}.
\label{eq:omega}
\end{equation}
For COMPASS, this radiative pion scattering reaction is then equivalent to
$\gamma$ +
$\pi$ $\rightarrow$ $\gamma$ + $\pi$ scattering for
laboratory $\gamma$'s with energy of order $\omega =1 GeV$ incident on a
target $\pi$
at rest.  
The function ${F}_{\gamma\pi}^{pt}{(}{\theta}{)}$ describing the Thomson cross
section for $\gamma$ scattering from a point pion is given by:
\begin{equation}
\label{eq:Primakoff_3}
{F}_{\gamma\pi}^{pt}{(}{\theta}{)}=
\frac{1}{2}\cdot
\frac
{1+{\cos}^{2}{\theta}}
{{(}{1+\frac{\omega}{m_{\pi}}{(1-\cos{\theta})}}{)}^2}
{.}
\end{equation}

From Eq.~\ref{eq:Primakoff_2}, the cross section depends on
$(\bar{\alpha}_{\pi}+\bar{\beta}_{\pi})$ at small $\theta$, and on
$(\bar{\alpha}_{\pi}-\bar{\beta}_{\pi})$ at large $\theta$. A precise fit of the
theoretical cross section (Eq.~\ref{eq:Primakoff_1}-\ref{eq:Primakoff_3}) to the
measured angular distribution of scattered $\gamma$'s, allows one to extract the
pion electric and magnetic polarizabilities. Fits will be done for different
regions of $\omega$ for better understanding of the systematic uncertainties. We
will carry out analyses with and without the dispersion sum rule constraint
\cite {hols1} that $\bar{\alpha}_{\pi}+\bar{\beta}_{\pi}\approx0.4$. We can
achieve a significantly smaller uncertainty for the polarizability by including
this constraint in the fits.

The event generator produces events in the alab frame, characterized by the
kinematic variables t, $\omega$ and $\cos\theta$, and distributed with the 
probability of the theoretical Compton Primakoff cross section
(Eq.~\ref{eq:Primakoff_1}-\ref{eq:Primakoff_3}). Then, the $\gamma\pi$ scattering
kinematics are calculated. The virtual photon incident along the recoil direction
$\vec{k}/|\vec{k}|$, is scattered on the pion "target", and emerges as a real photon
with energy/momentum $\omega^{\prime}$/$|\vec{k}^{\prime}|$ at an angle $\theta$:
\begin{equation}
\label{eq:Compton}
{\large
\omega^{\prime}=\frac
{\omega{(}{1}+\frac
{\omega^2{-}{|\vec{k}|}^2}{{2}{m}_{\pi}{\omega}}
{)}}
{
{1}{+}
\frac{\omega}{{m}_{\pi}}
{(}{1-}
\frac{|\vec{k}|}{\omega}
\cos\theta
{)}
}
}
\end{equation}

The photon azimuthal angle around the recoil direction is randomly generated
using a uniform distribution. The four-vector components of all reaction
participants (pion, photon and recoil nucleus) are then calculated in the alab
frame. The azimuthal angle of the recoil nucleus is also randomly generated by a
uniform distribution. Finally, the reaction kinematics are transformed to the lab
frame by a Lorentz boost.

For the measurement of the pion polarizabilities, one must fit the theoretical cross
section (\ref{eq:Primakoff_1}-\ref{eq:Primakoff_3}) to measured distributions, after
correcting for acceptances. The sensitivity to the polarizability increases with
increasing $\omega$ energy and at back angles. A convenient method is to use the
$\cos\theta$ distribution integrated over t and 
for different ranges of $\omega$, since this shows clearly
the sensitivity to the polarizability. 

\subsection{Design of the Primakoff Trigger\label{sec:how}}

The small Primakoff cross section and the high statistics required for extracting
polarizabilities require a data run at high beam intensities and with good
acceptance. This sets the main requirements for the trigger system: (1) to act as a
"beam killer" by accepting only Primakoff scattered pions, and suppressing the
high rate background associated with non-interacting
beam pions, (2) to avoid cutting the   acceptance at the important $\gamma$ back
angles in the alab frame, where the hadron polarizability measurement is most
sensitive, (3) to cope with background in the $\gamma$ calorimeter from low energy
$\gamma$'s or delta electrons.

We achieve these objectives via a COMPASS Primakoff trigger that makes use of a
beam veto, a target recoil detector, the calorimeters and various hodoscopes.  
A coincidence is required of the scattered pion with a $\gamma$ measured in the
ECAL2 calorimeter.  We demonstrated the feasibility and field operation of such
a trigger, via Monte Carlo simulations \cite {cd,bormio,tm,sans,kuhn,mlc} 
and via
beam tests
with the COMPASS spectrometer \cite{pvcs}.

For the reaction given in Eq. ~\ref{eq:polariz}, for illustration at 300 GeV pion
beam energy,
the laboratory outgoing $\gamma$'s are emitted within an angular cone of within
5 mrad, and the corresponding outgoing $\pi$'s are emitted within 2 mrad. Most
events have $\gamma$ energies between $0-280$ GeV, and $\pi$ energies between
$20-300$ GeV.

Our MC shows that we lose very little polarizability information by applying an
"energy cut" trigger condition that rejects events with the outgoing pion energy
having more than 240 GeV. Correspondingly, the final state $\gamma$ has less
than 60 GeV. The 240 GeV cut acts as a beam killer.  The 60 GeV cut will also be
very effective in reducing the $\gamma$ detector (ECAL2) trigger rate, since a
large part of the background $\gamma$ rate is below 60 GeV. 

\begin{figure}[tbc]
\centerline{\epsfig{file=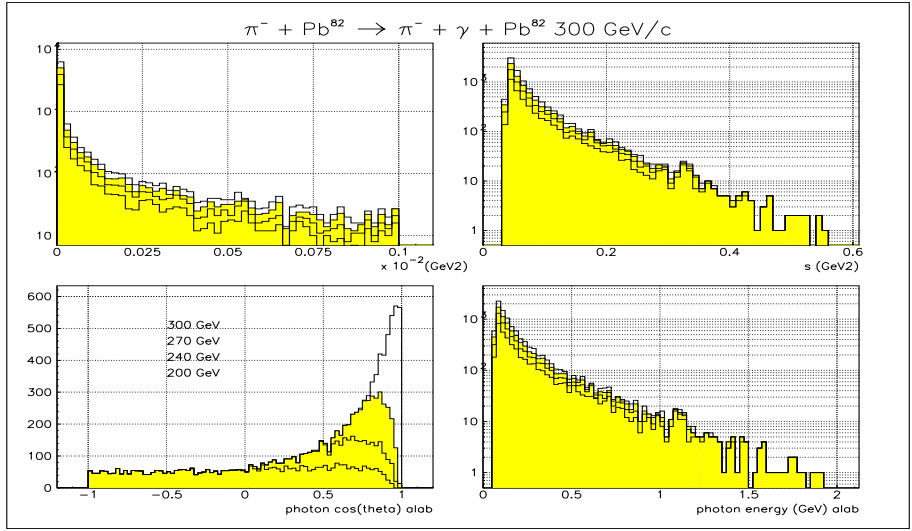,width=12cm,
height=7cm}}
\caption{MC simulation showing the acceptance of the
$\gamma\pi\rightarrow\gamma\pi$ reaction in terms of the invariant four
momentum transfer t to the target,the squared invariant energy s of the final
state $\gamma\pi$, the angular distribution versus $\cos(\theta)$ with
$\theta$ the $\gamma$ scattering angle in the alab frame, and the virtual photon
energy $\omega$ 
in the alab frame. The overlayed spectra correspond to different cuts
on the final state $\pi$ momentum.}
\label{fig:acceptance}
\end{figure}

The polarizability insensitivity to these cuts results from the fact that the
most forward (in alab frame) Compton scattering angles have the lowest
laboratory $\gamma$ energies and largest laboratory angles. In addition, the
cross section in this forward alab angle range is much less sensitive to the
polarizabilities. This is seen from Eq. \ref{eq:Primakoff_2}, since with
$\bar{\alpha}_{\pi}+\bar{\beta}_{\pi}\approx 1$ used in our MC, the
polarizability component is small at forward compared to back angles. The
acceptance is reduced by the energy cut for the forward alab angles (shown in
Fig.~\ref{fig:acceptance}), but is unaffected at the important alab back angles.
Summarizing, the pion and $\gamma$ energy constraints at the trigger level
fulfill the "beam killer" requirement and at the same time remove backgrounds
coming from low energy $\gamma$'s, delta electrons, and e$^+$e$^-$ pairs
incident on ECAL2, etc. Similar results are obtained for the effectiveness of an
energy loss trigger for simulations carried out at 190 GeV pion beam energy.

\begin{figure}[h]
        \centerline{\includegraphics[scale=0.8]{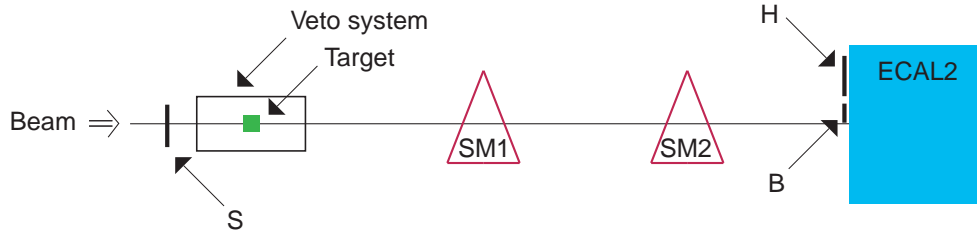}}
        \sl\caption{\small Trigger setup for the Primakoff measurement in 
COMPASS. 
	It consists of 
	a beam counter (S), beam veto counter (B), a hodoscope (H), a veto system around
	the target and the electromagnetic calorimeter (ECAL2).}
    \protect\label{trigsetup}
\end{figure}

Test beam studies for the trigger were performed in Sept. 2000~ \cite{pvcs}. 
The setup for the test beam was the following (see Fig. ~\ref{trigsetup}): 
a beam counter upstream of the 
target (S); a beam veto counter (beam killer) in front of ECAL2, covering the 
hole for the deflected primary beam (B); a hodoscope 80\cm\ $\times$ 96\cm , 
situated in front of ECAL2 (H), displaced horizontally by 20\cm\ from the 
position of the deflected beam; a veto system around the target and the 
electromagnetic calorimeter (ECAL2). 

 According to the signature of the 
reaction, a beam particle is expected in S. 
Particles scattered by the target, should give a signal in H. 
No signal should be registered in the beam killer B, but a highly energetic
photon 
should be detected in ECAL2. 
The test beam demonstrated that an acceptable rate for the trigger can be achieved. 
The requirements to achieve this rate are the following: the coincidence between the 
energy deposition in the ECAL2 above a threshold of 
0.2 to 0.3$\times E_{beam}$ and the 
existence of a charged particle in the corresponding acceptance of H. 
The use of the beam killer B and of the veto counter does not improve much the rate 
reduction. The target veto could be used offline 
to reject background reactions with large momentum transfer to the target. 

\begin{figure}[h]
\centerline{\epsfig{file=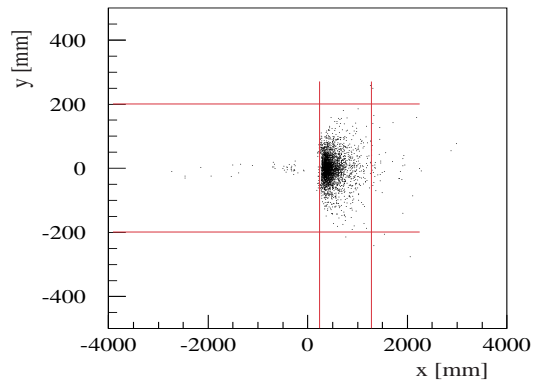,width=7cm,height=5cm}}
        \sl\caption{\small Simulated hit distribution 
of the scattered pion at 30 meters 
			from the target. The lines indicate the possible size of
			the trigger hodoscope.}
    \protect\label{trighod}
\end{figure}

Some interesting numbers that are quoted include the following:
from a beam intensity of $6\cdot 10^{6}$/spill with a 3\mm\ lead target, 
the trigger rate was found to be 2.5 $\times 10^5$/spill. The trigger
gives a reduction factor of 
24. The idea of the trigger is not to reduce too strongly the number of events 
since this will be done offline when analysing the events.
The size of the hodoscope needed to cover the acceptance of the simulated 
scattered pions was calculated by plotting the simulated hit distribution of the 
scattered pions at the position where the hodoscope will be situated. 
Fig.~\ref{trighod} shows that the size needed for the hodoscope is 1
meter 
in horizontal direction and 40 cm. in vertical direction. 

\begin{figure}[tbc] 
\centerline{\epsfig{file=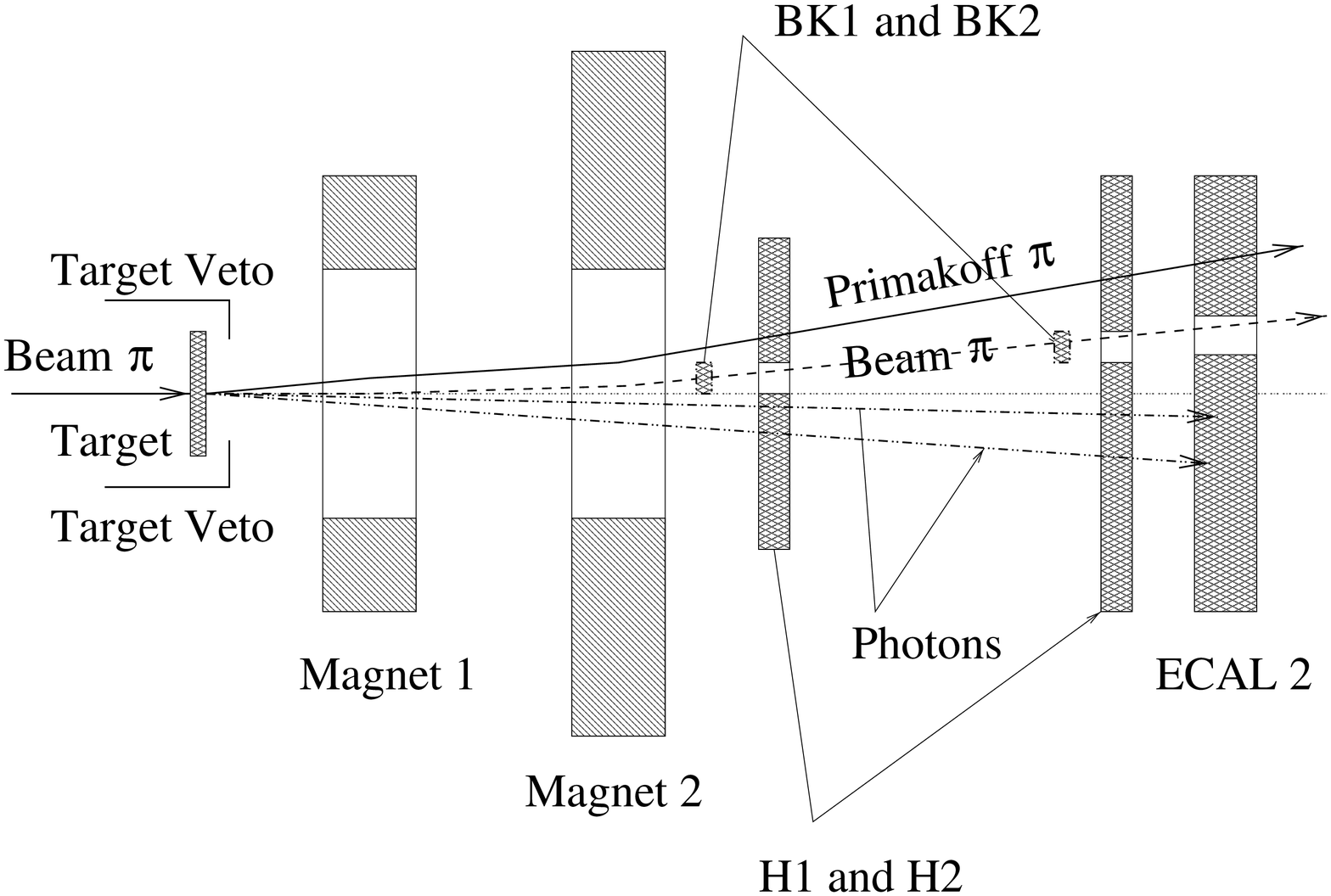,width=16cm,height=11cm}}
\caption{Detector layout for the COMPASS Primakoff Hybrid trigger,
$\pi^- Z \rightarrow Hybrid \rightarrow \pi^- \eta$.
BK1,BK2=beam killer system, H1,H2=hodoscope system for charged particle 
detection, ECAL2=second photon calorimeter.
For $\eta$ decay, one observes the 
two $\gamma$s shown.        For polarizability, there is only one 
$\gamma$ to 
detect.}
\label{fig:trigger}
\end{figure}

  Aside from the polarizability trigger discussed above, 
COMPASS will implement as well 
a meson radiative transition trigger.  
The trigger uses the
characteristic decay pattern: one or three charged mesons with gamma hits,
or three charged mesons and no gamma hits.  The trigger \cite
{bormio,tm} is based on a
determination of the pion energy loss (via its characteristic angular
deflection), correlated in downstream scintillator hodoscopes stations (H1
versus H2)  with the aid of a fast matrix chip, as shown in
Fig.~\ref{fig:trigger}. This trigger is a copy of the currently running 
muon beam energy loss trigger \cite {future}. 
We will use 
the Beam Kill veto trigger detectors BK1/BK2 only for low intensity tests. They 
follow
the pion trajectory, as shown in Fig.~\ref{fig:trigger}, but they can not handle 
the full 100 MHz beam rate. 

\subsection{Beam Requirements}

In COMPASS, two beam Cherenkov detectors (CEDAR) far upstream of the target
provide $\pi/K/p$ particle identification (PID). The incoming hadron momentum
is measured in the beam spectrometer. Before and after the target, charged
particles are tracked by high resolution silicon tracking detectors. The
measurement of both initial and final state momenta provides constraints to
identify the reaction. The final state pion and $\gamma$ momenta are
measured downstream in the magnetic spectrometer and in the $\gamma$
calorimeter, respectively.  These measurements allow a precise determination
of the small p$_T$ kick to the target nucleus, the main signature of the
Primakoff process, and the means to separate Primakoff from
meson  exchange
scattering events.

We can get quality statistics for the pion study with high beam intensities at
the CERN SPS.  We will take data with different beam energies and
targets, with
both positive and negative beams, as part of efforts to control systematic
errors. We will also take data with protons (positive beam) and 
antiprotons (negative beam with PID) to determine 
the proton 
polarizability \cite {an} by proton Primakoff
Compton
scattering, to self-calibrate our Primakoff Compton scattering
methodologies.

  For the 120-300 GeV hadron beams, particle identification (PID)
is needed to provide pion, kaon, proton beam tag for positive
and negative beams at different beam energies.  
For the
COMPASS beam, one expects \cite {beam}
a beam intensity of 100 
MHz, with
beam composition roughly: 120-300 GeV/c, negative, 87-98\% pions, 7-1\% kaons,
2\%-1\% antiprotons;  120-300 GeV/c, positive, 43-2\% pions, 7-1\% kaons,
49-97\% protons.
PID is accomplished at CERN with CEDAR
detectors, CErenkov Differential counter with Achromatic Ring focus.  
There are two CEDAR detectors (in series) in the COMPASS
beamline \cite {bov}.
They each have eight large area PMTs arranged in a circle,
preceded by a single light diaphragm (LD) to finely fix a ring radius.  
For the PID, a six-fold coincidence is required. The gas pressure is
varied to set the ring radius of pions or Kaons or protons at the LD
location, to choose one of the particle types.  The Narrow diaphragm
mounted in CEDAR-N separated kaons from pions up to 300 GeV/c and could
detect protons down to 12 GeV/c.

\subsection{Target and Target Detectors\label{sec:target}}

The target platform is moveable and allows easy insertion of a normal solid
state target, e.\,g.\ a cylindric lead plate 40\mm\ in diameter and 3\mm\ in
width. We also plan Primakoff scattering on nuclei with Z$<82$ to check the
expected Z$^2$ cross section dependence, and to allow corrections 
for double photon exchange \cite {ak}. We use 
silicon tracking detectors
before and immediately after the targets.  These are essential for Primakoff
reactions as the scattering angle has to be measured with a precision
of order 100 $\mu$rad, because it contains much of the
information.  We veto target
break-up events by a target veto recoil detector, and by selecting low-t
events in the off-line analysis. The recoil detector is currently developed
and tested. A particle momentum of about 100\mev\ should be sufficient to
trigger a veto coincidence in its scintillator and Lead glass layers. This
veto will limit the angular acceptance to about 100\mrad\ to reject events
where fragments of the nucleus or particles produced in diffractive processes
leave the target with bigger angles to the beam axis.

\subsection{The Magnetic Spectrometer and the t-Resolution}

We achieve good momentum resolution for the incident and final state pions and
$\gamma$'s, and therefore good four momentum t-resolution.  The relative
momentum resolution for \Ppim\ with all interactions accounted for is 1\% for
energies above 35\gev\ and up to 2.5\% below this mark. The angular resolution
in a single coordinate for a pion of momentum p is 7.9 mrad-Gev/p. All
generated pions with energy greater than 2\gev\ were in the acceptance of the
spectrometer, the reconstruction efficiency with interactions enabled is 92\%.

The angular resolution for the final state $\pi$ is controlled by minimizing the
multiple scattering in the targets and detectors. With a lead target of 1\%
interaction length (2 g/cm$^2$,~30\% radiation length), multiple Coulomb
scattering (MCS) of the beam and outgoing pion in the target gives an rms
angular resolution of order 40 $\mu$rad, small compared to the intrinsic
tracking detector angular resolution. The target contributes to the resolution
of the transverse momentum p$_T$ via the p$_T$ generated through MCS.  
Considering $t~=~p_T^2$, including all other effects \cite{cd,bormio,tm}, 
we
aim for a p$_T$ resolution less than 15 MeV, corresponding to $\Delta t$ better
than $\approx$ 2.5 $\times 10^{-4}$ GeV$^2$.

This resolution will allow an effective t-cut to minimize contributions to the
data from diffractive processes. The goal is achievable, based on the t
distributions measured at a 200 GeV low statistics, high resolution experiment
for $\pi^- Z \rightarrow \pi^- \pi^0 Z$ \cite{jens} and $\pi^- Z \rightarrow
\pi^-
\gamma Z$ \cite{ziel} Primakoff scattering at 200 GeV at FNAL. The t
distribution of the $\pi^- \rightarrow \pi^- \gamma$ data agrees well with the
Primakoff formalism out to t~=~$10^{-3}$ GeV$^2$, which indicates that the
data are indeed dominated by Coulomb production. Minimum material (radiation
and interaction lengths) in COMPASS will also give a higher acceptance, since
that allows $\gamma$'s to arrive at ECAL2 with minimum interaction losses, and
minimum $e^+e^-$ backgrounds.

\subsection{The $\gamma$ Calorimeter ECAL2}

The position resolution in the second $\gamma$ calorimeter ECAL2 for the photon
is 1.0\mm\ corresponding to an angular resolution of 30\murad. In the
interesting energy range the energy resolution is better than 1\% after taking
into account the leakage into the hadron calorimeter HCAL. The photon acceptance
is 98\% due to the beam hole of ECAL2 while the reconstruction efficiency is
58\% as a result of pair production within the spectrometer.

This $\gamma$ detector is equipped with 3.8 by 3.8 cm$^2$ GAMS lead glass blocks
to make a total active area of order 1.5 m diameter.  The central area needed
for the polarizability measurement is 
completely instrumented with modern ADC readouts.

The p$_T$ kicks of the COMPASS magnets are 0.45 GeV/c for SM1 (4 meters from
target) and 1.2 GeV/c for SM2 (16 meters from target).  The fields of both
magnets are set $\it{additive}$ for maximum deflection of the beam from the zero
degree (neutral ray) line.  We thereby attain at least 10 cm for the distance
between the zero degree line and the hole edge. This is important since the
Primakoff $\gamma$'s are concentrated around the zero degree line and a good
$\gamma$ measurement requires clean signals from 9 blocks, centered on the hit
block.  From MC simulations, the number of Primakoff scattered pions below 40
GeV is less than 0.3\%, so that 40 GeV pions are about the lowest energy of
interest.

For the precise monitoring of the energy calibration of the photon calorimeters,
COMPASS will use LED and laser monitor systems, as described in Ref. 
\cite{laser}. 

The available COMPASS ECAL does not have radiation hardened blocks near the 
beam hole. If those will be available, it would allow increasing the beam 
intensity a factor of five, to allow substantially more statistics for the same 
run time. 

\section {Simulation Results for 190  GeV Pion Beam}

 The most recent COMPASS Primakoff polarizability simulation results 
\cite {sans,kuhn,mlc}
are given now.
We evaluated the single particle detection properties of
the \compass\
spectrometer. The key results are
a momentum resolution for pions of 1\% above 35\gev\ and up to 2.5\%
below 35\gev\ accompanied by an angular resolution of $7.9\mrad\gev/p$
and an energy resolution for photons of 1.5\% above 90\gev\ 
accompanied by an angular resolution of 30\murad. 
The photon
reconstruction inefficiency is given by the conversion probability
before leaving the second spectrometer magnet; the corresponding
efficiency is 58\%. The single pion reconstruction efficiency is about
92\%.

We describe the results of the simulation of Primakoff Compton scattering and of
hadronic backgrounds. We study the influence of the COMPASS detector resolutions
and reconstruction algorithms on the extraction of the polarizabilities.  We
achieve this via a simulation of Primakoff Compton scattering of \Ppim\ on Lead
at a beam energy of 190\gev\ at five different pairs of
$\bar\alpha$ and
$\bar\beta$ 
polarizability parameters with a total statistics of five times 620,000
events, each
sample corresponding to the analyzed events from roughly a day of \compass\ data
taking.

The following cuts are used to recognize Primakoff events: there has
to be a photon hit in the ECAL2 above a certain energy and a negative
charged track which carries the complementary energy. The cut on the
photon energy has to be well above the energy deposition of hadrons in
the electromagnetic calorimeter that is of the order of some\gev. In
order to select the so-called ``hard events'' with most information on
the pion polarizabilities, this cut is raised to 30--50\% of the beam
energy.  In this simulation a cut on the photon energy at 90\gev\ was
implemented in the generator, so it was natural to use 80\gev\ in the
reconstruction. The energy window for the sum of pion and photon was
set to 180--196\gev\ to allow for the longitudinal energy leakage of
ECAL2. The figures 6-9 were all produced only with the events that were
left after applying these two cuts.

The reconstructed count rates have to be corrected for the inefficiency of the
detector before fitting the cross section.  The data of all five samples are
merged to calculate the dependence of the reconstruction efficiency on the
photon energy in the laboratory frame. \Figref{sim:rec:photon} shows the
generated and---with the cuts applied---the reconstructed photon energy in the
laboratory frame, \Figref{sim:rec:eff} shows the corresponding efficiency with
fit parameters. As expected from our studies of single photon and pion
efficiencies, the overall efficiency is between 50-55\%. Fig. 8 shows the
beam energy distribution used in the simulation.

Compared to the generated $t=\lvec q^2$ distribution,
the reconstructed $\lvec q^2$ is smeared out 
due to the transverse momentum transfer error induced
by angle reconstruction errors of the final state.  For example, quadratically
adding the errors for 110\gev\ photon energy and 80\gev\ pion momentum yields an
error of 15\mev. 
Nevertheless the rise is steep enough to permit a cut at
$-\lvec q^2<1000\mev^2$, as seen in Fig. 9. Such a cut only reduces the
efficiency by 6\%.

 We studied the efficiency for the selected events with small $t$ and where the
sum of the energy of the $\pi^-$ and the $\gamma$ is, within some resolution,
equal to the energy of the beam.  We found that this efficiency was independent
of $t$, with no bias by the acceptance of the detector. Regardless of the cut in
$t$, the important shape of the $\gamma \pi$ angular distribution is not
affected.

\twoplots{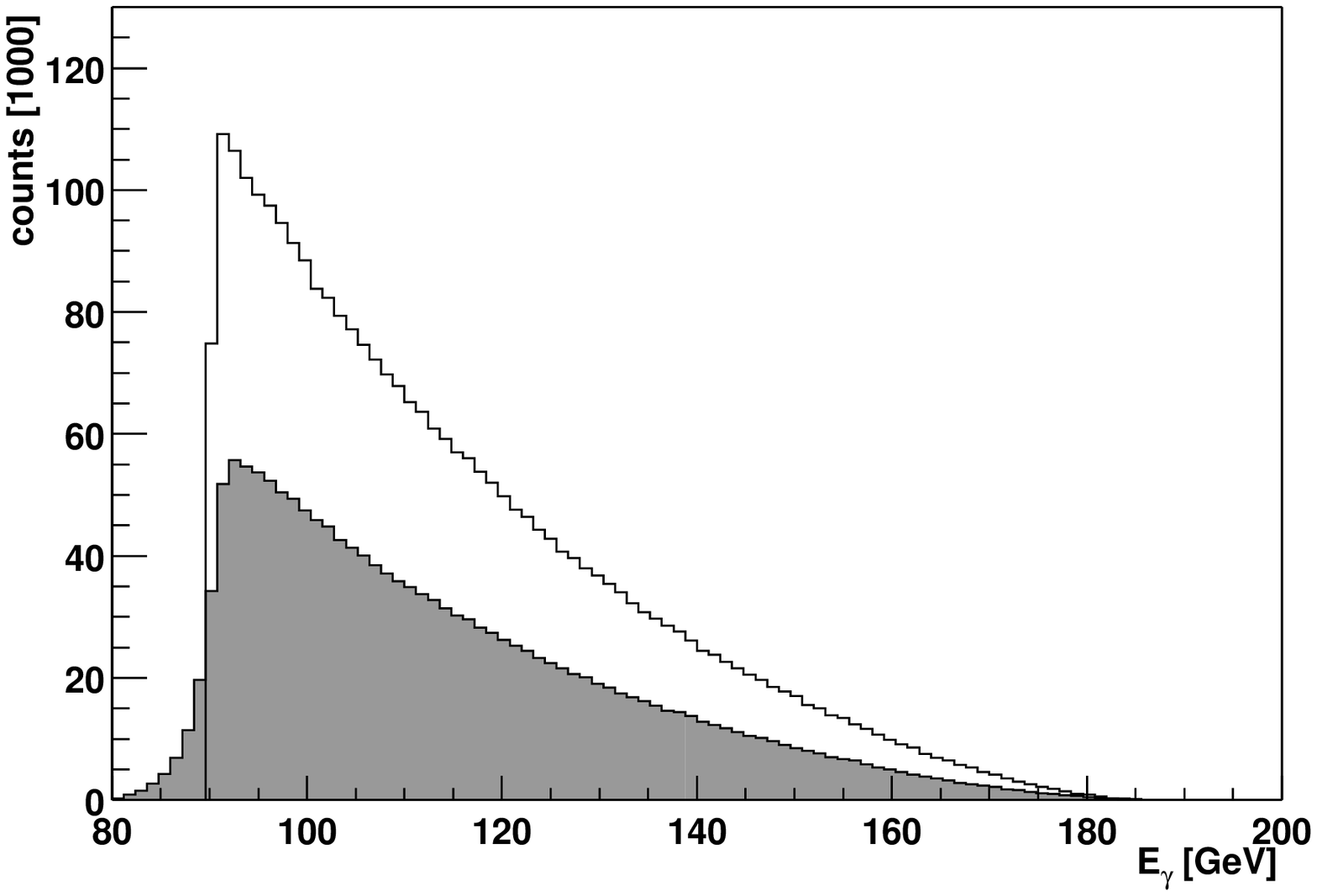}{Photon energy in the laboratory
  frame}{generated (white) and re\-con\-struc\-ted (shaded) photon
  energy in the laboratory frame. The data of all five samples is
  merged ($2.9\cdot10^6$ events)}{sim:rec:photon}
{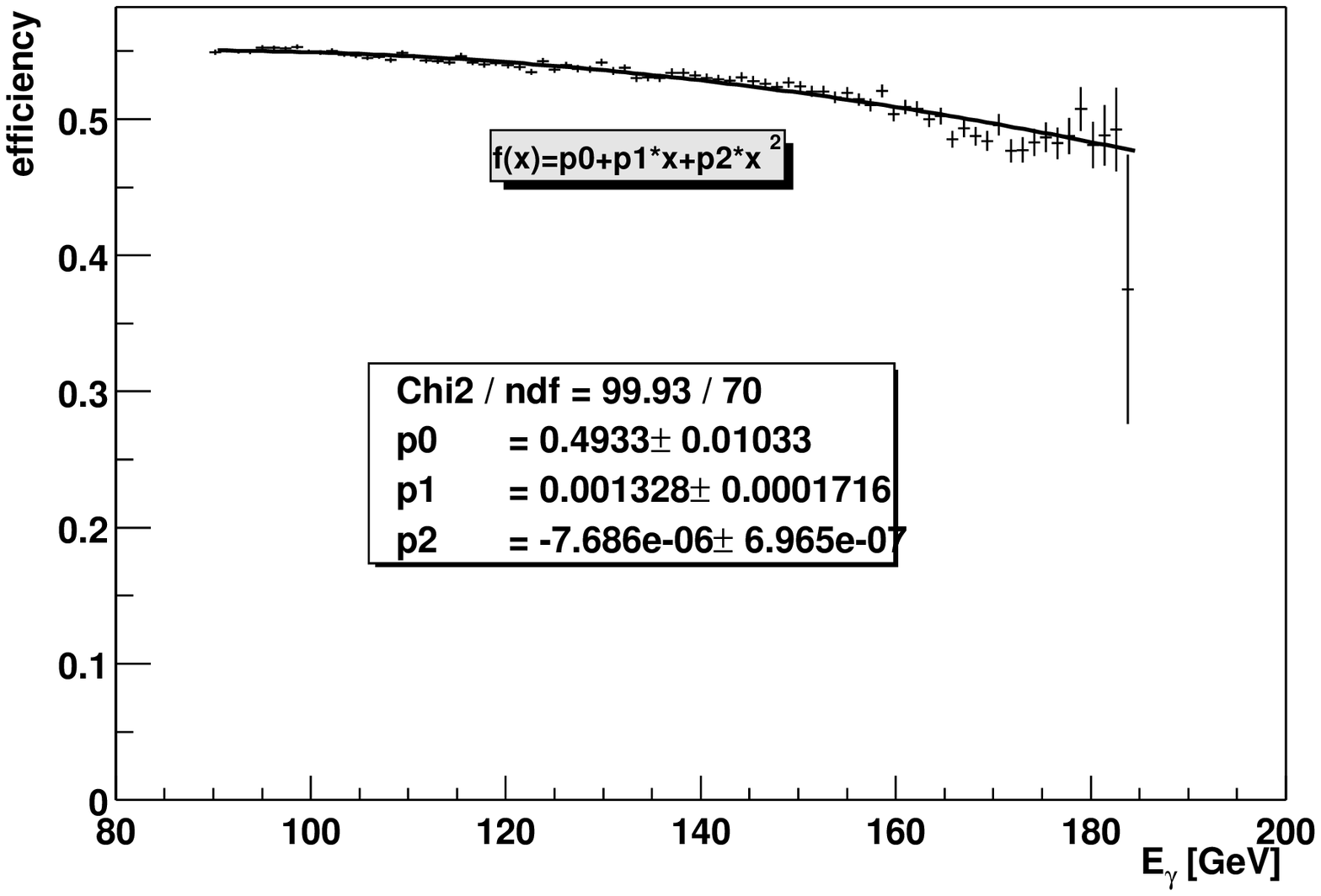}{Reconstruction efficiency for
  Primakoff}{reconstruction efficiency vs. photon energy in the
  laboratory frame. The error bars are the binomial errors
  corresponding to the generated statistics shown in the left hand
  plot}{sim:rec:eff}

\twoplots{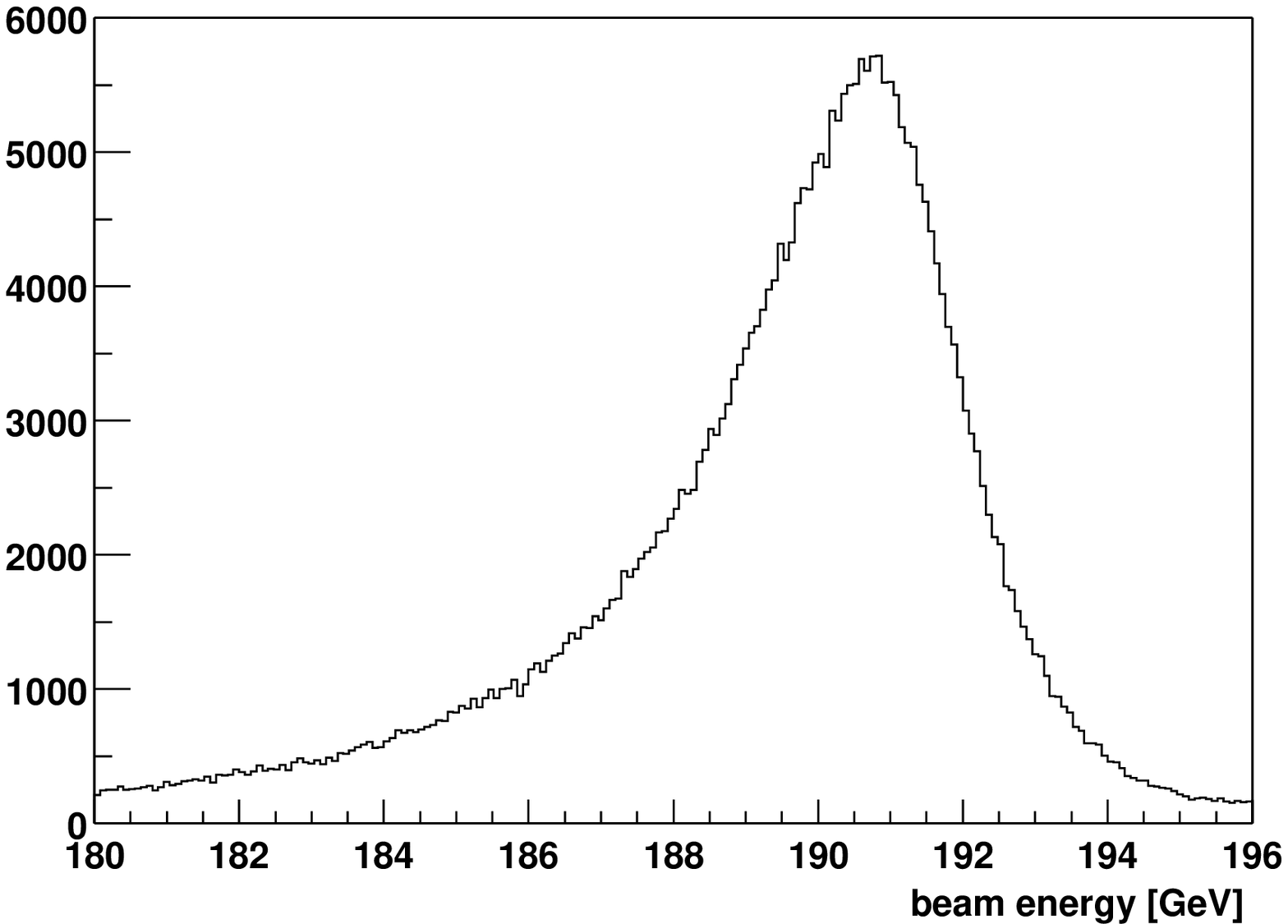}{Reconstructed beam
  energy}{reconstructed beam energy}
{sim:rec:beam}{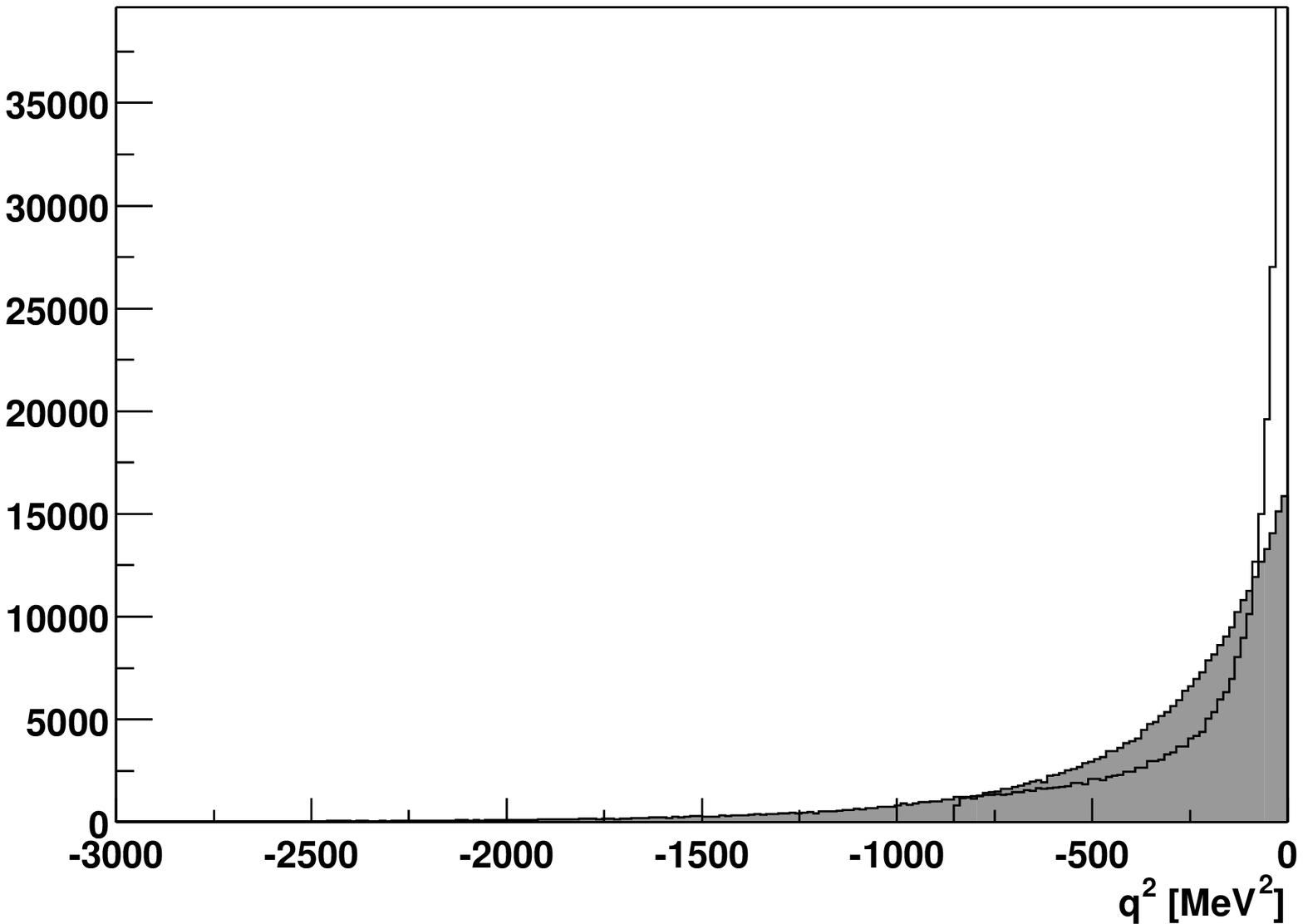}{Four-momentum transfer: $\lvec
  q^2$}{generated (white) and re\-con\-struc\-ted (shaded) $\lvec
  q^2$}{sim:rec:q2}

\subsection{Retrieving the polarizabilities}

The generated and reconstructed
four-momenta are transformed to the projectile (alab) frame. 
The fit is performed using the \ROOT\ interface to the MINUIT
package. 
To get the reconstructed differential cross section the event rates
had to be corrected for the inefficiency of the detector.
We observe that $\oalpha$ and $\obeta$ are anti-correlated, with the consequence
that  
$\oalpha+\obeta$ is determined with a much smaller error than
$\oalpha-\obeta$. The polarizability
effect is proportional to $\oalpha+\obeta$ for $\cos\theta=1$ and
$\oalpha-\obeta$ for $\cos\theta=-1$, increasing with the photon
energy like $\omega^2$. So $\oalpha-\obeta$ is mostly determined in
the region with least events while $\oalpha+\obeta$ is better known
because the cross section has a steep rise towards positive
$\cos\theta$.

  We carried out fits to determine the polarizabilities for each sample of
620,000 simulated events, each sample corresponding to different 
($\oalpha$,$\obeta$) pairs. Reducing the statistical errors by 8, scaling to
the
assumed  
4. $\times 10^{7}$ events, we estimate 
the statistical uncertainties for the two month COMPASS run to be about 0.05
for $\oalpha$ and $\obeta$, 0.01 for
$\oalpha+\obeta$ and 0.1 for $\oalpha-\obeta$. Including systematic
uncertainties, we aim to achieve better than 0.4 uncertainties in
$\oalpha$ and $\obeta$.

\subsection{The hadronic background}

To investigate the corruption of the measured Primakoff cross section
by hadronic background events, a large sample of minimum bias events
was produced with the Fritiof pion-Nucleus event generator. 
The analysis of $4.5 \times 10^6$  events by the exact
process described above accepts only 34 events, 27 of them were in the
fit range for the polarizabilities \cite{kuhn}.

As Fritiof only simulates hadronic interaction the mechanism for accepting
some of the events is the production of \Ppiz\ or \Peta. Because of the cut
on the total energy sum of the pion and the photon, most of this background is
rejected, as the remaining particles also receive their part of the energy.
One may tighten this cut because the background would be much more affected
than the real events.

The generated final state in all 34 cases contains a nucleon. This suggests
that the nucleus was disintegrated in the reaction. The fragments are tracked
by \comgeant, but there is no single particle ID to label them. Thus, they
unfortunately are not part of the list of particles emerging from the primary
vertex. Every event contains particles with polar angles bigger than
20\degree\, and a target recoil would see some of those events not stopped
inside the target.

The overall signal to noise ratio for hadronic background can be estimated from
the ratio of the cross sections and the background suppression. The hadronic
interaction length of Lead of $194\g\cm^{-2}$ corresponds to a cross section of
1.77\b, the suppression factor of 26/4500000 reduces this to 10\mub. This has to
be compared to the cross section of Primakoff Compton scattering---with a
produced photon energy of at least 90\gev---of about 500\mub. The ratio of 50:1
will be further improved by the inclusion of the target recoil veto and by a more
sophisticated kinematics reconstruction.

\section{Other Pion Polarizability Experiments}

  The present proposal deals only with our experimental efforts at CERN COMPASS.   
But we will also compare COMPASS data to new Mainz data.  At 
MAMI-B at Mainz,
measurements \cite {mainz} and calculations \cite {un00} are under way of
$p(\gamma,n \pi^+ \gamma')$ radiative pion photoproduction reaction on the
proton. The elastic $\gamma \pi^+$ scattering cross section can be found by
extrapolating such data to the pion pole. This corresponds to "Compton"
scattering of gamma's from virtual $\pi^+$ targets in the proton, and therefore
also allows a measure of the pion polarizability. The experiment ran with
500-800 MeV tagged photons, with detection of $\gamma'$, neutron, and charged
pion in coincidence. A similar pion polarizability experiment with polarized
tagged photons was proposed \cite {hi99}, associated with the proposed JLab 12
GeV upgrade. Theoretical studies \cite {un00} show that the precision of this
method, aside from the available statistics,  is limited by the 
systematic errors associated 
with the extrapolation to the pion
pole. Primakoff scattering has the advantages 
that meson exchange backgrounds and final state interactions are strongly
suppressed.  Pion polarizabilities have also been determined from
two photon $\gamma \gamma \rightarrow \pi^+ \pi^-$ data \cite {babu2}.
However, as noted in the pion polarizability theory review of  Portoles and 
Pennington \cite{Pennington}: "Knowing the low 
energy  $\gamma \gamma \rightarrow \pi^+ \pi^-$  cross section accurately still 
allows large uncertainties in the polarizabilities. Only measurements of Compton 
scattering will resolve these." 
Still, intercomparisons between COMPASS and past plus new
experiments \cite {babu2,mainz,hi99}, with complementary methodologies, 
should 
help fix the
systematic uncertainties.

\section{Conclusions}

COMPASS
is on track to measure the $\gamma\pi$ Compton 
scattering cross sections, as a central part of its Primakoff
physics program, thereby enabling determinations of the pion 
polarizabilities. The experiment will allow serious tests of $\chi$PT;
and of different available polarizability calculations in
QCD.

\section {Acknowledgments} 
 
This research was supported in part by the Israel Science Foundation founded by
the Israel
Academy of Sciences and Humanities.


\end{document}